\documentclass{conm-p-l}
\usepackage[dvips]{graphics}
\usepackage{epsf}

\copyrightinfo{2003}{J\"urgen Fuchs, Ingo Runkel and Christoph Schweigert}

\theoremstyle{definition}

\theoremstyle{remark}

\numberwithin{equation}{section}
\def\be            {\begin{equation}}

\def\calc          {{\mathcal C}}
\def\calg          {{\mathcal G}}
\def\calh          {{\mathcal H}}

\def\calm          {{\mathcal M}}

\def\complex       {{\mathbb C}}
\def\eE            {{\rm e}}

\def\eps           {\varepsilon}
\def\eq            {\,{=}\,}
\newcommand\erf[1] {(\ref{#1})}
\def\F             {{\mathbb F}}
\newcommand\Frac[2]{\mbox{\large$\frac{#1}{#2}$}}
\def\Hom           {{\rm Hom}}
\newcommand\hsp[1] {\mbox{\hspace{#1 em}}}
\def\id            {{\rm id}}
\def\ii            {{\rm i}}
\def\iN            {\,{\in}\,}
\newcommand\labl[1]{\label{#1}}
\def\Lie           {{\rm Lie}\,}
\newcommand\nxt[1] {\\\raisebox{.12em}{\rule{.35em}{.35em}}\hsp{.6}#1}
\def\one           {{\bf 1}}
\def\oti           {\,{\otimes}\,}
\def\reals         {{\mathbb R}}
\def\RI            {Riemann}

\def\scs           {\scriptstyle }

\def\U             {{\rm U}}
\def\zet           {{\mathbb Z}}

\begin{document}

\begin{flushright} \mbox{$\,$}\\[-13mm]
                             {\tt HU-EP-03/02}\\[1mm] {\tt hep-th/0301181}
\\[22mm]{\,}\end{flushright}

\title [Lie algebras, Fuchsian differential equations and CFT correlators]
       {Lie algebras, Fuchsian differential equations
       \\ and CFT correlation functions}

\author{J\"urgen Fuchs}
\address{Avdelning fysik, Karlstads universitet,
Universitetsgatan 5, S\,--\,651\,88\, Karlstad}
\email{jfuchs@fuchs.tekn.kau.se}
\author{Ingo Runkel}
\address{ Institut f\"ur Physik, HU Berlin,
Invalidenstr.\ 110, \,D\,--\,10115 Berlin}
\email{ingo@physik.hu-berlin.de}
\author{Christoph Schweigert}
\address{Institut f\"ur theor.\ Physik,
RWTH Aachen, Sommerfeldstr.\ 14, D\,--\,52074\, Aachen}
\email{schweige@physik.rwth-aachen.de}

\date{December 31, 2002}
\dedicatory{}


\begin{abstract}
Affine Kac-Moody algebras give rise to interesting systems of differential 
equations, so-called
Knizhnik-Zamolodchikov equations. The monodromy properties of their solutions
can be encoded in the structure of a mo\-du\-lar tensor category on 
(a subcategory
of) the representation category of the affine Lie algebra. We discuss the 
relation between these solutions and physical correlation functions
in two-dimensional conformal field theory.
In particular we report on a proof for the existence of the latter on 
world sheets of arbitrary topology. 
\end{abstract}

\maketitle

\section{Some venerable differential equations}

One of the surprises in the theory of Kac-Moody algebras is the observation
that they give a new and powerful handle on problems that, a priori, do not
have an algebraic flavor, including such which are not even genuinely 
infinite-dimensional. In this contribution we discuss one such application of
structures related to affine Kac-Moody algebras: properties of solutions to 
differential equations.

\medskip

In the middle of the 18th century Euler, Cauchy and other analysts, 
the successors of the founding fathers like 
Newton, Leibniz, and Johann and Jakob Bernoulli, developped 
a strong interest in ``special functions'' and series. Arguably, the 
simplest non-trivial series one can think of is the geometric series
  $$ f(z) = 1+z + z^2 +\ldots  $$
It obeys the ordinary differential equation
  $$ z\,(1\,{-}\,z)\, f'' + (1\,{-}\,3z)\, f' - f = 0 \, . $$
To be of interest in complex analysis, a generalization of the geometric series 
should, of course, possess a non-zero radius of convergence. Furthermore,
a useful guiding principle is that it should also obey a simple differential 
equation that allows one to determine its main properties. Around the year 
1750, such considerations lead Euler to the {\em hypergeometric series}
  $$ F(\alpha,\beta,\gamma;z):= 
  1 + \sum_{n=1}^\infty  \frac {\alpha(\alpha{+}1)\cdots (\alpha{+}n{-}1)
  \,\beta(\beta{+}1)\cdots(\beta{+}n{-}1)} {n!\, \gamma(\gamma{+}1)\cdots
  \,(\gamma{+}n{-}1)}\, z^n \,.$$
It obeys the hypergeometric differential equation
  \begin{equation}
  z(z\,{-}\,1)\, w'' + [(\alpha\,{+}\,\beta\,{+}\,1)\,z -\gamma]\,w' 
  +\alpha\beta\, w = 0 \,. \labl{dg} \end{equation}
This linear differential equation is Fuchsian \cite{fuch},
i.e.\ enjoys the particularly nice property that
all three singular points that the coefficient functions acquire upon 
division by $z(z{-}1)$ (namely 0, 1, and $\infty$) are regular singularities.
This means that when one approaches
such a point in a wedge region of opening angle less than $\pi/2$ the 
solutions will not diverge stronger than polynomially.

It is quite remarkable that Euler's interest in this series did {\em not\/} 
arise from concrete applications. Indeed, in his famous 1857 paper \cite{riem}
Riemann calls Euler's motivation a ``theoretisches Interesse'', while
he points out there are ``numerous applications in physical and astronomic
investigations''. (This makes a strong point in favour of
pure science that is motivated by internal questions.)

Differential equations of the type \erf{dg} and their solutions are
indeed ubiquitous in mathematical physics. They include
\nxt Legendre polynomials, introduced by Legendre around 1800 in his
     study of gravitational potentials:
  $$ \frac1{\sqrt{1-2\rho z + \rho^2}} = \sum_{n=0}^\infty \rho^n P_n(z)
  \,; $$
\nxt Bessel functions, describing the radial part of the wave function of
     a free particle in quantum mechanics;
\nxt Laguerre polynomials, the radial part of the hydrogen atom wave
     functions.
\\
The last two classes of functions are {\em confluent\/} hypergeometric
functions; they are defined by the limit
  $$ \Phi(\alpha,\gamma;z ) = \lim_{\beta\to\infty} 
  F(\alpha,\beta,\gamma;\Frac z\beta) \, . $$

So far there is clearly not much algebra in the game. A first hint on the
possible relevance of algebra comes from the observation that
a deeper understanding of many special functions is afforded by relating 
them to representation functions of suitable
Lie groups. For example, Bessel functions are related to representation
functions of irreducible unitary representations of the group of
motions of the Euclidean plane. (For a review of the relation 
between special functions and Lie groups, see e.g.\ \cite{HElg,DIeu}.)

In this contribution, we consider interesting generalizations of
hypergeometric differential equations that have their origin in 
representation-theoretic structures, involving in particular the 
representation theory of (classes of) Kac-Moody algebras. 
They are largely motivated by models of conformal quantum field theory.

In Section 2 we introduce a system
of differential equations. In section 3 properties of their solutions
are discussed, in particular the asymptotic behaviour
and monodromies. These properties possess a convenient interpretation
in terms of tensor categories. In section 4 we finally give 
the underlying motivation from theoretical physics and present
a theorem about correlation functions in conformal field theories.

\section{The Kniznik-Zamolodchikov equation}

Let us select a real compact Lie group $G$. It admits a bi-invariant metric
that allows us to introduce dual bases $\{a_l\}$ and $\{a^l\}$ of the Lie
algebra $\Lie G$ of $G$. The element
  $$ \Omega := \sum_{l=1}^{\dim\,G} a_l \oti a^l \,\in \Lie G \oti \Lie G $$ 
does not depend on the choice of bases.
Moreover, its action on the tensor product $V_1\oti V_2$ of two 
finite-dimensional $G$-modules $V_1$ and $V_2$ is diagonalizable.

We are interested in functions 
  \be f: \quad \calm_N \to V_1\otimes \cdots \otimes V_N  \labl7
  \end{equation} 
on the moduli space 
  $$ \calm_N:= \complex^{\times N}\setminus \Delta  $$ 
of $N$ mutually distinct points in the complex plane with values in the 
tensor product of $N$ finite-dimensional $G$-modules.
For such functions we consider the system 
  \be  D_i\,f(z_1,z_2,...\,,z_N) = 0 \,, \qquad i\eq1,2,...\,,N\,,  \labl1
  \end{equation}
of differential equations, with differential operators 
  \be  D_i := \kappa\,\partial_{z_i} - \sum_{\scs j=1 \atop\scs i\neq j}^N
  \frac{\Omega_{ij}}{z_i-z_j} \, . \labl{con}
  \end{equation}
Here the symbol $\Omega_{ij}$ stands for $\Omega$ acting on the tensor
product of the $i$th and $j$th modules, i.e.
  $$  \Omega_{ij} = \sum_l \one \oti\cdots\oti \one \oti R_i(a_l) 
  \oti\one \oti\cdots\oti \one \oti R_j(a^l)
  \oti\one \oti\cdots\oti \one $$
with the non-trivial factors at the $i$th and $j$th positions.
Further, $\kappa$ is a real parameter; for the moment we assume that 
it is irrational, but eventually, in the context of conformal quantum 
field theories, we will be particularly interested in certain rational 
values of $\kappa$. 

Let us mention one particular physical system in which the differential 
equation \erf1 arises: a spin chain. Take $G$ to be $SU(2)$
and, for all $i\eq1,2,...\,,N$, $V_i$ the two-dimensional defining 
representation of $SU(2)$. This system models a one-dimensional metal --
a chain of $N$ metal atoms -- with each atom having just
one electron, with spin described by $V_i$, in the outermost shell. The 
tensor product in \erf7 is then the space of physical states. 
For this spin chain one considers a Hamiltonian
  $$ H(u) = \sum_{\scs i,j=1 \atop\scs i\neq j}^N \frac{\Omega_{ij}}
  {(u-z_i)\,(z_i-z_j)} + \sum_{i=1}^N \frac{\calc_i}{(u-z_i)^2} $$
depending on parameters $(z_1,z_2,...\,,z_N)\iN\calm_N$
and $u\iN\complex\setminus\{z_1,z_2,...\,,z_N\}$.
The first term in $H$ describes a hopping from position $i$ to position $j$,
while the second term, involving the quadratic Casimir $\calc_i$, is
an internal energy that depends on the element of the chain.

One aims at diagonalizing the Hamiltonians $H(u)$ for all values of $u$
simultaneously. One approach to this problem, the so-called Bethe ansatz 
technique (see e.g.\
\cite{GAud}), leads to \erf1 as an auxiliary problem \cite{bafl,feFr}. Other
motivations to study \erf1 comes from low-dimensional quantum field theory,
knot theory and, somewhat surprisingly, the theory of modular representations 
of finite groups of Lie type. These aspects will be addressed later on.

\section{Properties of the solutions}

We now discuss some properties of the solutions to \erf1. We first study how
they behave when one approaches singular points of the coefficients and
then have a look at their monodromies. Our discussion in this section follows 
the exposition in \cite{BAki}.

\subsection{Asymptotic properties}

Two properties of the solutions to \erf1 are immediate:
\nxt By summing over $i$ and using that $\Omega_{ji}\eq\Omega_{ij}$, one obtains
  $$ \sum_{i=1}^N \partial_{z_i} f(z_1,z_2,\ldots,z_N) = 0 \,.  $$
Thus the solutions are invariant under simultaneous 
translation of all arguments,
  $$ f(z_1{+}\zeta,z_2{+}\zeta,...\,,z_N{+}\zeta) = f(z_1,z_2,...\,,z_N) \,. $$
\nxt Second, one finds that 
  $$ \kappa \sum_{i=1}^N z_i\, \partial_{z_i} f
  = \sum_{\scs i,j=1 \atop\scs i<j}^N \Omega_{ij} f \,.  $$
Hence the solutions to \erf1 have a well-defined scaling behaviour:
  $$ f(\lambda z_1,\lambda z_2,...\,,\lambda z_N) = 
  \lambda^{\kappa^{-1} \sum_{i<j} \Omega_{ij}} f(z_1,z_2,\ldots,z_N) \,. $$

\medskip

For $N\eq3$, these two properties allow us to restrict our attention to the
situation that $z_1\eq0$, $z_2\eq t$ and $z_3\eq1$. We can therefore simplify 
the problem and consider functions
  $$ f:\quad (0,1) \to V_1\otimes V_2\otimes V_3 $$
that obey the ordinary differential equation
  \be \kappa\, f'(t) = \mbox{\LARGE(} \frac{\Omega_{12}}t
  + \frac{\Omega_{23}}{t-1} \mbox{\LARGE)}  f(t) \, . \labl2
  \end{equation}
This is a Fuchsian differential equation of the type presented in the
first section.

Next we study the asymptotics of the solutions $\Gamma(V_1,V_2,V_3)$ to 
the equations \erf2 on the open interval $(0,1)$ of the real line,
a problem analogous to the one studied by Frobenius in \cite{frob}. 
Restricting $t$ to be real, we have chosen a well-defined way of approaching
the two singular points $t\eq 0,1$ of the coefficient functions. Let
$v\iN V_1\oti V_2\oti V_3$ be an eigenvector of $\Omega_{12}$,
  $$ \Omega_{12} v = \lambda\, v \, . $$
If $\kappa$ is irrational, then
there is exactly one solution behaving in the limit $t\,{\to}\, 0$ as
  $$ t^{\lambda/\kappa}\, (\,v +
  \mbox{\small vector valued analytic function}\,) \,. $$
Therefore the limit $t\to 0$ provides us with a bijection
  $$ \Phi_0 :\quad \Gamma(V_1,V_2,V_3) \to V_1\oti V_2\oti V_3 \,.  $$
One can show 
that $\Gamma(V_1,V_2,V_3)$ carries the structure of a $G$-module, and that
the bijection $\Phi_0$ is compatible with the $G$-action and hence constitutes
an intertwiner of $G$-modules. (In particular, $\Phi_0$ restricts to an
isomorphism between the respective subspaces of $G$-invariants.) Similarly,
studying the solution close to $t\eq1$ with the help of eigenvectors of 
$\Omega_{23}$ one obtains a different identification
  $$ \Phi_1 :\quad \Gamma(V_1,V_2,V_3) \to V_1\oti V_2\oti V_3 \,.$$
Combining the two maps we obtain a $G$-module intertwiner
  $$ \alpha_{V_1,V_2,V_3} = \Phi_1^{}\Phi_0^{-1} :\quad
  (V_1\oti V_2) \oti V_3 \to V_1 \oti (V_2\oti V_3) \,. $$

We can now introduce a new non-strict tensor product of $G$-modules by taking 
$\alpha$ as the associator that defines the rule for changing brackets in 
a two-fold tensor product. In short, considering different asymptotics 
of solutions to differential equations allows us to 
define a different notion of associativity, and thus a different tensor
product, on the category of $G$-modules. It turns out that this modified 
tensor product is the relevant one for the conformal field theories in 
which the differential equations \erf1 arise, and in particular that it fits 
with the infinite-dimensional algebraic structures that we will encounter
in that context below.  
  
One can now proceed to study different limits of solutions to equation 
\erf1 on the open simplex 
  $$ z_1 < z_2 < \cdots < z_N $$
to define multiple tensor products. One verifies the consistency
of different ways of changing brackets. (Technically, this amounts to
showing the pentagon axiom for the associator; for more details we refer to
\cite{BAki}.)

\subsection{Monodromy}

A different way to understand properties of the solutions to \erf1 is to 
analyze the differential operators \erf{con}.
They mutually commute, $ [D_i , D_j]\eq 0 $, and therefore define a flat
connection on a trivial vector bundle over the moduli space $\calm_N$. 
Analytic continuation of the solutions to \erf1 around singular points 
yields their monodromies.
The monodromies generate a representation of the fundamental group 
$\pi_1(\calm_N)$. (Studying the solutions via their monodromies is
the point of view of Riemann \cite{riem}.)
One can also derive integral representations of the solutions, generalizing
those familiar from the hypergoemetric function, which provide a tool for 
calculating the monodromies; see \cite{VArc} and references cited there. 

We can use the flat connection to exchange insertion points, as illustrated 
in the following picture:
  \be  \begin{array}l \begin{picture}(210,50)(0,30)
  \put(44,0)  {\begin{picture}(0,0)(0,0)
              \scalebox{.15}{\includegraphics{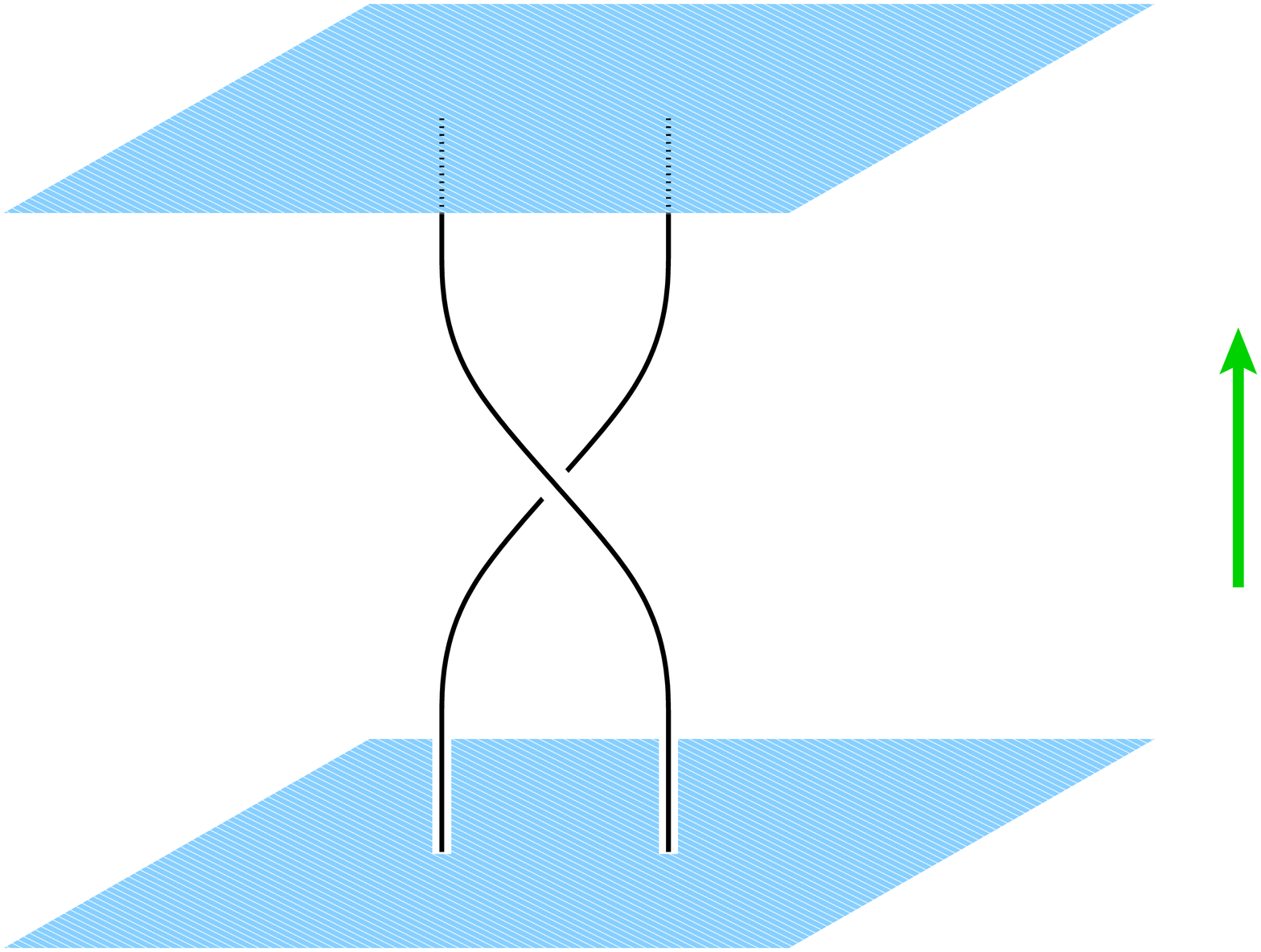}} \end{picture}}
  \put(68.5,48)   {\scriptsize$z_i(t)$}
  \put(103.1,48)  {\scriptsize$z_j(t)$}
  \put(158.5,40)  {$t$}
  \end{picture}\\{}\\[1.9em]\end{array} \labl5  \end{equation}
Suppose we have selected paths $z_i(t)$ for exchanging the points. Then 
our task is to integrate the ordinary differential equations
  $$ \kappa\, f' = \sum_{i<j} 
  \frac{ z_i'-z_j'}{z_i-z_j}\, \Omega_{ij} f \,, $$
where the prime indicates derivative with respect to $t$.
This gives us a new rule 
  $$ c_{V_1,V_2}^{} := P_{12} \eE^{\pi\ii \Omega/\kappa}:
  \quad V_1\oti V_2 \to V_2\oti V_1  $$
for exchanging vectors. This operation
does not square to the identity map; as a consequence, what one represents
on $V_1\oti V_2$ is no longer the permutation group. With some good will, 
in the picture \erf5 one recognizes braids -- and indeed, the fundamental
group $\pi_1(\calm_N)$ is the pure braid group on $N$ strands.

\section{Why is all that interesting?}

\subsection{Infinite-dimensional structures}

To explain the interest of our problem, let us sketch three different
applications. 
     The first two will be discussed in the present subsection, the third in
     subsection 4.4.  
The algebraic structures to which we have related the 
analytical problem that we started with have so far been finite-dimensional.
But they are actually closely related to infinite-dimensional structures.

\smallskip

(1) The picture \erf5 is reminiscent of knots. Using the fact that
   the trivial representation is contained in the tensor product of a
   representation and its dual, one can close the braids and thereby
   obtain knot invariants and, more generally, invariants of links in 
   three-manifolds. In particular, one obtains Jones' invariant 
   for $G\eq SU(2)$. 
   This invariant first appeared \cite{jone2} in the study of inclusions of 
   von Neumann algebras. The latter are infinite-dimensional algebras -- this
   is a first indication that infinite-dimensional structures are hidden 
   in the problem under study. As it turns out, most of the analytic structure
   of von Neumann algebras is indeed not essential for this problem.  

   The hypergeometric equation discussed above (multiplied with suitable 
   powers of $t$ and $1{-}t$) is e.g.\ obtained when considering the
   invariants in the tensor product of two defining (highest weight
   $\Lambda_{(1)}$) and two conjugate defining (highest weight $\Lambda_{(n-1)}$)
   representations of $G\eq SU(n)$. (Concretely, the relevant hypergeometric
   functions are
   ${{}^{}_{2\scriptscriptstyle\!}}F_1^{}(\alpha,-\alpha,1{-}\beta;t)$ and
   ${{}^{}_{2\scriptscriptstyle\!}}F_1^{}(\alpha{+}\beta,\alpha{-}\beta,
   1{+}\beta;t)$ with
   $\alpha\eq\frac1{n+k}$ and $\beta\eq\frac n{n+k}$, where $k\iN\zet_{>0}$
   \cite{knza}.) To give another example, for the invariants in the tensor
   product of four defining representations of $G\eq SO(n)$ the solutions 
   can be expressed through double contour integrals that are very similar,
   though different from, the integral representation of the generalized
   hypergeometric functions ${{}^{}_{3\scriptscriptstyle\!}}F_2^{}$,
   with parameters (appearing as exponents in the integral representation)
   that are again simple combinations of $\frac1{n+k}$ and
   $\frac n{n+k}$ with $k\iN\zet_{>0}$ \cite{jf7}.

   Unfortunately, it seems that this approach to knot theory has not
   completely lived up to its original hope. Indeed, while the invariants
   constructed from the monodromies of the solutions can separate
   knots in a much finer way than classical invariants (they can, for instance,
   distinguish between a knot and
   its mirror image), it is still unknown whether they can separate all
   non-isomorphic knots. Also, to our knowledge these techniques have so 
   far not really lead to a proof of any classical conjecture about knots
   or the topology of three-manifolds.

\smallskip

(2) 
The second motivation to study solutions to the equations \erf1 comes from
infinite-dimensional algebras which arise in conformal quantum field
theories in two dimensions. Such theories have many applications:
\nxt In string theory, the notion of a point-particle is replaced by a
     one-dimensional object, called a string. Instead of a world line, one
     deals with a two-dimensional world sheet, which is embedded into
     space-time. After imposing the so-called conformal gauge condition,
     regarding the space-time coordinates as fields on the world sheet
     $X$ gives rise to a conformal field theory on $X$.
\nxt They describe universality classes of two-dimensional critical
     phenomena. Recently, the theory of critical percolation has received 
     particular attention (for a review, see \cite{card19}).
\nxt In condensed matter theory, chiral conformal field theory has found
     applications in the description of (universality classes of) quantum Hall
     fluids. Conformal field theory on surfaces with boundary has proven to
     be useful to understand the properties of
     point-like defects like in the Kondo effect.
\nxt Moreover, the study of conformal field theories has given rise to
     much interaction between physicists and mathematicians. This interaction
     has been particularly intense in the theory of infinite-dimensional
     Lie algebras and, more specifically, of Kac-Moody algebras.
     But various other mathematical disciplines have been involved as well.

\subsection{Correlation functions in quantum field theory}

In the setting of point (2) above,
a quantum field theory consists, schematically speaking, of two pieces of data:
\\[.2em]
i) A collection $\calh_i$ of spaces of physical states, also called
   {\em superselection sectors\/}. 
   The theory of infinite-dimensional Lie algebras, in particular of
   affine Kac-Moody algebras, plays an important role in the construction
   of interesting classes of quantum field theories. In the affine Kac-Moody
   case, the superselection sectors are obtained from the objects in
   in the category of integrable modules of finite length and of central
   charge  $\kappa{-}h(G)$, where $h(G)$ is the dual Coxeter number of $G$.
   A toy model that already displays many features of these (sub)categories
   is the category of finite-dimensional modules of a compact Lie group $G$.
\\[.2em]
ii) A collection of {\em correlation functions\/}. 
    Let us, for simplicity, consider a quantum field theory defined on the
    complex plane. Then for every $N\iN\zet_{\ge0}$ the existence of a 
    function on the moduli space $\calm_N$ of $N$ distinct points in the plane
    is required for every $N$-tuple of states $v_i\iN\calh_i$.  
    These functions, called ($N$-point) {\em correlation functions\/} or
    {\em correlators\/}, will be denoted by
  $$ {\bf C}_{v_1,\ldots,v_N}: \quad  \calm_N \to \complex \,.$$
The physical interpretation of ${\bf C}_{v_1,\ldots,v_N}$ is that the expression
  $$ {\bf C}_{v_1,\ldots,v_N}(z_1,z_2,...\,,z_N)=
  \langle\, \Phi(v_1;z_1)\cdots \Phi(v_N;z_N) \,\rangle $$
should be regarded as the correlation function of ``fields''
$\Phi(v_i;\cdot\,)$ that are located at the ``insertion points'' $z_i$.
Thus a quantum field theory can be thought of
as an infinite collection of functions on the moduli spaces $\calm_N$. 

The symmetries of a quantum field theory are implemented on the correlation 
functions by differential equations known as {\em Ward identities\/}.
In the conformal quantum field theories that are formulated with the help of
affine Kac-Moody algebras (known as WZW theories), the Ward identities include 
in particular a system of equations that are of the form of the differential 
equations \erf1.  It is worth mentioning that the correlation functions obtained 
as solutions to these equations are {\em exact\/}, i.e.\ do not involve any
approximation or expansion whose convergence is not under control.
In particular there is no need to resort to any kind of perturbation theory 
-- a virtue that results from formulating the theory in algebraic terms.

However, the correlation functions must not only respect the symmetries of the
theory, but are in addition subject to several other consistency constraints. 
For example, the theory must possess so-called cluster properties, i.e.
  \be \begin{array}{r}
  \displaystyle\lim_{\lambda\to\infty}\!\lambda^{\sum_{i=1}^k\! \Delta_i}_{}\, 
  \langle\, \Phi(v_1{;}\lambda z_1)\cdots \Phi(v_k{;}\lambda z_k)\,
  \Phi(v_{k+1}{;}z_{k+1})\cdots \Phi(v_N{;}z_N) \,\rangle
  \\{}\\[-.8em]
  = \, 
  \langle\, \Phi(v_1{;} z_1)\cdots \Phi(v_k{;} z_k)\,\rangle \,\,
  \langle\, \Phi(v_{k+1}{;}z_{k+1})\cdots \Phi(v_N{;}z_N) \,\rangle
  \end{array}\!\!\! \labl{clus} \end{equation}
for all $1\,{<}\,k\,{<}\,N$, with $\Delta_i$ the conformal weight, or scaling 
dimension, of the field $\Phi(v_i;\cdot\,)$. These constraints express the fact 
that the interactions of the theory are local in the sense that if we take groups 
of fields very far apart, their mutual interference can be neglected, so that
asymptotically, in the limit of large distances, the correlation function 
approaches the product of correlators with less insertions.

\subsection{Modular tensor categories}

To present a third reason for the interest in equation \erf1,
a bit more background information is needed.
The differential equation \erf1 allows us to endow the category
$\calc(G)$ of finite-dimensional $G$-modules with the structure of
a braided tensor category. This is a category with a tensor product 
(typically non-strict, i.e.\ with a non-trivial associator) and a 
braiding. A braiding is a collection of prescriptions 
  $$ c_{V,W} :\quad V\oti W \to W \oti V $$
for permuting modules that one obtains from the holonomy with 
respect to the flat connection \erf{con}. Moreover, the notion 
of conjugate modules endows this category with a duality.

One can show -- the proof being surprisingly hard \cite{drin7,drin9,kalu} 
-- that the category obtained this way is equivalent to the category
of representations of the quantum group $\U(\Lie G)_q$ at a generic
value of the deformation parameter $q$.

Interesting quantum field theories that are realized in two-dimensional
physical systems do indeed lead to such categories. Having both braiding 
and duality morphisms, the categories obtained from these systems also 
possess a twist; it is related to the conformal weight $\Delta$
of the superselection sectors. In the following we restrict our attention to 
tensor categories which in addition are {\em modular\/}, i.e.\ 
possess two additional properties:
\\
i)~~~There are only finitely many isomorphism classes of simple objects $U_i$.
\\
ii) The square matrix $s\eq(s_{ij})$ that is obtained by taking the 
trace of twice the braiding of simple objects, 
  $$  \begin{array}l \begin{picture}(30,27)(0,27)
  \put(44,0)  {\begin{picture}(0,0)(0,0)
              \scalebox{.38}{\includegraphics{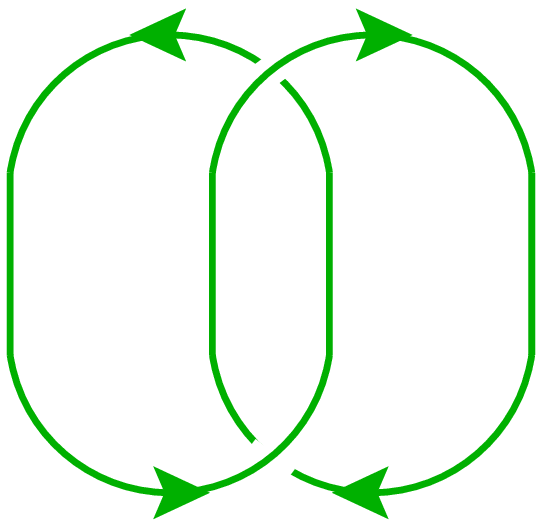}} \end{picture}}
  \put(-69,25.3)  {$s_{ij} \,:=\, {\rm tr}\,c_{U_i,U_j}^{}c_{U_j,U_i}^{} \,=$}
  \put(57.4,22.8) {\scriptsize$U_i$}
  \put(81.9,22.8) {\scriptsize$U_j$}
  \end{picture}\\{}\\[1em]\end{array} $$
is non-degenerate.

Concerning the finiteness of the number of (classes of) simple objects,
some comments are in order:
\nxt
An example for an algebraic structure with a finite number of irreducible
representations are finite groups. Indeed, each finite group gives rise to
a modular tensor
category, although not in a completely naive way. The reason is, roughly
speaking, that Fourier transformation can be defined naively only for
functions on abelian groups. There is, however, also a notion of
``exotic Fourier transform'' \cite{lusz16}. These categories appear
in so-called orbifold conformal field theories \cite{dvvv,bant4,dolm5}.
\nxt 
Generic values of the deformation parameter $q$
of the quantum group $\U(\Lie G)_q$ correspond to irrational values
of the parameter $\kappa$ that is present in the differential operators
\erf{con}. When $\kappa$ is chosen to be rational, a truncation to finitely many 
(isomorphism classes of) simple objects is also possible. (In this case,
logarithms can appear in the asymptotics of the solutions to \erf1. These
usually do not contribute to the special solutions that in
addition satisfy the locality condition ({\bf Loc}) that will be
discussed below.) An interesting range of values for $\kappa$ are the 
integers larger than the dual Coxeter number $h(G)$ of $\Lie G$, 
which leads to unitary theories.
The simple objects of the truncated category then correspond to 
the level $\kappa{-}h(G)$ integrable irreducible highest weight 
modules of the untwisted affine Kac-Moody algebra based on $\Lie G$.
For the representation category of the quantum group $\U_q(\Lie G)$,
an analogous truncation is possible when $q$ is a root of unity;
this truncation is closely connected to the truncation in the 
Lie algebra case, with the relation between the deformation parameter  
and $\kappa$ given by $q\eq\eE^{2\pi\ii/\kappa}$.

\subsection{Modular representations} 

Now we are in a position to present the third motivation for the study
of the differential equations \erf1: modular representations of finite 
groups like $SL(\F_p)$ (with $\F_p$ the field with $p$ elements) on vector 
spaces over fields of the same characteristic $p$. It is amazing that such
a purely algebraic problem requires (complex-)\,analytic tools as well as 
structures arising in connection with infinite-dimensional Lie algebras. 
But the study of such representations is difficult indeed, and even their 
dimension was unknown ten years ago.

The central idea is that in finite characteristic a similar truncation
happens as upon deformation. One therefore relates representation theoretic
problems -- most importantly, the problem of describing the composition series 
of the highest weight module obtained via the Borel-Bott-Weil construction --
to problems for quantum groups. Their representation category,
or rather a certain truncation thereof, carries the structure of a 
modular tensor category. One shows that it is (anti-)equivalent to a
category of modules over affine Lie algebras which obtains its structure
of a modular tensor category precisely from \erf1. In the latter category,
it is comparatively easy to work with character formulae and related structures. 
(For a review, see \cite{ande2} and the last section of \cite{math7}.)

\medskip

For completeness, we finally mention that to some extent the logic we 
presented can be reversed. That is, each modular category gives rise to 
monodromy data, in fact for arbitrarily many insertion points on a Riemann 
surface of any genus. However, one and the same modular tensor category 
(and hence monodromy data) can appear as the representation category of 
more than one conformal field theory.
Indeed one even knows of examples where infinitely many
conformal field theories possess the same modular tensor category: those 
based on the groups $SO(n)$ and level 1, with fixed value of $n\bmod 16$.

\section{Existence of correlation functions}

\subsection{Complexification}

The schematic summary of (two-dimensional conformal) quantum field 
theories presented above is, of course, an oversimplification. In 
particular, their structure is more complicated in the following respect. 
The world sheet -- the two-dimensional manifold $X$ on which the theory
is living -- need not be the complex plane, or equivalently the Riemann
sphere, but can be a Riemann surface of arbitrary genus. It may also have 
a boundary, as happens for the world sheet of open strings, and need not 
be orientable. Furthermore, to obtain well-understood differential
equations such as \erf1, one does not directly study the correlation functions 
on the world sheet $X$ that arises in the physical application, but 
(multi-valued) functions on a different two-dimensional manifold.

This situation is familar from potential theory. 
Harmonic functions on $\reals^2$ are either analytic in $z\eq x{+}\ii y$ 
or analytic in $\bar z\eq x{-}\ii y$. A convenient trick when
working with harmonic functions is to consider instead of functions on
$\reals^2$ functions on a complex $z$-plane and on a complex $\bar z$-plane.
This allows us to work in a complex setting. Note that that to
achieve this we must work with a {\em double\/} of the manifold.

This manipulation becomes more transparent when one considers the case
that $X$ has a boundary. Then the prescription, including again a doubling 
of the space, is well known from classical electrodynamics. Consider a point
charge at position $y\iN\reals^n$ in front of an infinite grounded 
hyperplane located at $x^n\eq0$.
Without boundary, the electrostatic potential obeys the Poisson equation
  $$ - \Delta_x V_{y}(x) =  q\, \delta(x-y) \,, $$
which is solved by the Greens function; in two dimensions,
  $$ V_y(x) = G(x,y) = - \Frac1{2\pi} \, \log\,|x-y| \,. $$
To include the effects of the grounded plane, one introduces a mirror charge
$-$ the position of which is given by reflection $\sigma$ at 
the plane. The potential in the presence of the plane then reads
  $$ V(x) = G(x,y) - G(x,\sigma y) $$
Notice that again the space we use for solving the problem is a double cover
of the original space on which the physical problem is formulated.
 
Further, in two-dimensional conformal field theory
it is natural \cite{scfu4} to endow the space $X$ with
the structure of a real scheme. The simple differential equations \erf1
-- the Ward identities --
and their solutions -- often called {\em conformal blocks\/} -- then live
on the complexification $\hat X$ of $X$, which is a complex scheme.
Let us give some examples. If $X$ does not have any real point (intuitively,
in the language of manifolds: if the boundary $\partial X$ is empty), then,
as a manifold, $\hat X$ is just the total space of the orientation bundle 
over $X$. In particular, when $X$ is orientable and $\partial X\eq\emptyset$, 
then the double $\hat X$ is the disjoint union of two copies of $X$
with opposite orientation, $\hat X\eq X \,{\sqcup}\,({-}X)$. Two more
examples: the complex double of the disk is the Riemann sphere, and the double
of an annulus is a torus. The double $\hat X$ carries
the action of an anticonformal involution $\sigma$, which is
nothing but the action of the Galois group
$\calg al(\complex,\reals)\,{\cong}\,\zet_2$ on the complexification. One
may identify $X$ with the quotient of $\hat X$ by this action,
  $$ X = \hat X / \{1,\sigma\}  \, . $$ 
The insertion points on $\hat X$ are the pre-images, under the natural
projection from $\hat X$ to $X$,
of the insertion points $p_i$ on $X$. Thus except when $p\iN\partial X$,
a field insertion $\Phi(v;p)$ on $X$ gives rise to two insertions on $\hat X$.

\subsection{Consistency conditions}

We have thus arrived at what is known as the {\em principle of holomorphic 
factorization\/}:
the physical correlation functions on $X$ are specific solutions to the
differential equations \erf1 on the double $\hat X$. These specific
solutions must obey the following additional consistency constraints: 

\noindent
\begin{itemize} \addtolength\itemsep{4pt}
\item[({\bf Loc})] 
They must be {\em local\/}, i.e.\ genuine functions on $X$; thus  
their monodromies must be trivial. This means that one looks for solutions 
to \erf1 on $\hat X$ that are invariant under the subgroup of 
the mapping class group of $\hat X$
that commutes with $\sigma$. This subgroup can be identified with the
mapping class group of $X$. This requirement is known as modular invariance;
it includes also moduli of the conformal structure on $X$. 
\item[({\bf Fac})] 
They must possess {\em factorization\/} properties, which generalize 
the cluster property \erf{clus}. 
\\[1mm]
(For a precise statement of factorization see, for instance, theorems
3.9 and 3.10 of  \cite{fffs3}.)
\end{itemize}

\vskip1mm

\noindent
A central problem for understanding conformal quantum field theories is 
to select among the conformal blocks on the double $\hat X$ -- 
solutions to differential equations with non-trivial monodromies -- 
a consistent collection of correlation functions. 
A priori it is far from clear whether this is possible at all. 
On the other hand, for specific classes of theories necessary conditions
on correlation functions have been discussed already long ago.
In particular, consequences of the locality condition ({\bf Loc}) for
correlation functions on surfaces of genus 0 have been discussed in
the form of associativity constraints on the operator product expansion, 
see e.g.\ \cite{dofa,knza}, and in the form of modular invariance 
constraints on the partition function (i.e., 0-point correlation 
function) on the torus (genus 1), the latter possessing always the 
so-called $C$-diagonal solution. However, at the time it was not known 
whether these solutions are part of a system of correlation functions 
on world sheets of arbitrary genus that solves all the constraints.

Only recently, a complete affirmative answer to the existence question 
could be given in the so-called Cardy case:
It was shown \cite{fffs3} that indeed the $C$-diagonal 
torus partition function is part of a system of correlators  satisfying
all constraints. The proof is constructive, describing every correlator 
as the invariant of a concrete ribbon graph in a three-manifold.
This description is possible due to the fact that the representation
category of the CFT -- in the present paper a full subcategory of the
representation category of an infinite-dimensional Lie algebra --
possesses the structure of a modular tensor category.

What is not settled by the result of \cite{fffs3} is the question of
uniqueness of the solution. For the case of the torus partition function
and for some very specific correlators at genus zero, solutions to
({\bf Loc}) that are not of the $C$-diagonal form were also already known 
for quite a while. In the latter examples the check of locality
involves application of rather special identities.
For instance, for certain 4-point correlators in WZW theories based on
the group $SU(n){\times}SU(n')$, for which the blocks are products 
of blocks for $SU(n)$ and for $SU(n')$, one finds non-$C$-diagonal
local combinations at the special values $k\eq n'$ and $k'\eq n$
of the levels, using e.g.\ \cite{jf8} a relation between products
of hypergeometric functions and generalized hypergeometric functions
as well as ${{}^{}_{3\scriptscriptstyle\!}}F_2^{}(\alpha,-\alpha,\frac12;
\beta,1{-}\beta;t) + \frac t{2nn'}{{}^{}_{4\scriptscriptstyle\!}}
F_3^{}(1{+}\alpha,1{-}\alpha,\frac32,1;1{+}\beta,2{-}\beta,2;t)\eq1$.
(Compare also appendix E of \cite{aflu4} for an application to the
Kondo effect.)

\subsection{The main theorem}

To find a collection of correlation functions that obey the two 
constraints ({\bf Loc}) and ({\bf Fac}) is a highly non-trivial problem.
Two strategies are a priori envisageable: the first one aims at reducing 
these constraints to a small set of necessary and sufficient conditions.
Due to the complexity of the problem, this approach has quite a few
pitfalls (see e.g.\ \cite{bakI}). In fact, the most popular way-out has been to
work with a set of necessary conditions and to hope that they are
sufficient as well. Here we adopt an opposite strategy
and give a simple {\em sufficient} condition that guarantees the existence of a
consistent system of correlation functions for a conformal field theory.
Our approach is based on a combination of the structure of braided tensor 
categories with concepts from non-commutative algebra. 
It leads to the following central result:

\smallskip

\noindent
\begin{itemize}\item[{}]
{\bf Theorem} \cite{fuRs}:\\
When the monodromies of a system of conformal blocks are described by
a modular tensor category, then a consistent set of correlation functions
solving ({\bf Loc}) and ({\bf Fac}) exists for each quadruple 
$(A,m,\eta,\eps)$ of the following data: $A$ is an object in the modular 
tensor category (and thus a superselection sector), $m$ is a multiplication
  $$ m:\quad A\oti A \to A $$
that is associative with respect to the associator, 
  \be  m \circ (m\oti \id_A) = m \circ (\id_A \oti m) \circ \alpha_{AAA}\,, 
  \labl{mult} \end{equation}
and $\eta\iN\Hom(\one,A)$ a unit morphism with respect to $m$.
\\
Finally, the morphism $\eps\iN\Hom(A,\one)$ is an appropriate generalization 
of a linear form that turns the algebra $(A,m,\eta)$ into a so-called 
{\em symmetric special Frobenius algebra\/}. 
\end{itemize}

\vskip1mm\noindent
(For precise definitions we refer to \cite{fuRs4}.)

\medskip

The proof of this statement has not yet been fully published. 
It uses heavily non-commutative algebra in tensor
categories. Here we can only make a few comments: \\
1) In order to turn the object $A$ into a symmetric special Frobenius algebra,
   only a single non-linear condition must be solved, namely the 
   associativity requirement \erf{mult}.
 \\[.2em]
2) The proof makes use of a three-manifold $M_X$ that is bounded by
   $\hat X$ and that has $X$ as a retract. Every correlation function can be
   described as the invariant of a ribbon graph in $M_X$, with the 
   building blocks of the graph corresponding to morphisms in the modular
   tensor category. (An important ingredient
   in the construction of the graph is a triangulation of $X$ \cite{fuRs}.)
   The consistency constraints ({\bf Loc}) and ({\bf Fac}) can then be 
   verified by manipulating ribbon graphs in $M_X$.
\\[.2em]
3) For various physical reasons one is interested in manifolds $X$ with
   non-empty boundary. Correlation functions on such $X$ depend on the choice
   of a ``boundary condition'' on $\partial X$. The treatment of this situation
   is possible owing to a simple
   description of boundary conditions: they correspond to modules
   (in the tensor category $\calc$) of the Frobenius algebra $A$. This allows us
   to apply representation theoretic methods \cite{fuSc16,ostr} to the study of
   boundary conditions in conformal field 
   theory, and thus e.g.\ to describe D-branes in certain string theoretic models
   on Calabi-Yau spaces.
   In a similar way, so-called ``defect lines'' correspond to $A$-$A$-bimodules.
\\[.2em]
4) It is not known so far (though we expect it to be the case) whether
   the existence of an associated quadruple $(A,m,\eta,\eps)$ is not only
   a sufficient, but also a necessary condition for a solution to the
   constraints ({\bf Loc}) and ({\bf Fac}) to exist, and in particular,
   whether {\em every\/} solution is obtained in the way described by the 
   theorem. In any case, 
   solutions and quadruples $(A,m,\eta,\eps)$ are {\em not\/} in bijection.
   Rather, Morita equivalent algebras give identical solutions. When one
   combines this insight with orbifold techniques, it allows one to give a
   rigorous proof of T-dualities and algebraic versions of mirror symmetry. 
\\[.2em]
5) The construction of \cite{fuRs} can be generalized in such a way that
   correlation functions can also be assigned to manifolds without orientation
   \cite{fuRsx}. In this case, the datum of a symmetric special Frobenius 
   algebra is not enough to determine the theory. Rather, one must in 
   addition require that the algebra $(A,m,\eta)$ is
   isomorphic to the opposite algebra $(A,m{\circ}c_{A,A},\eta)$. The choice
   of an algebra isomorphism, if it exists, enters as an additional datum.

\section{Conclusions}

Our starting point was the analytic problem of selecting consistent
correlation functions in a conformal field theory.
We have then realized that an algebraization of this problem allows us
to obtain general proofs of consistency.
Incidentally, we also gain considerable computational power. 
In particular, one can express what physicists call the ``structure 
constants of the operator product expansion'' in terms of invariants
of links in three-manifolds, which makes them explicitly computable 
in concrete models and allows one to prove some of their properties. 
Another virtue of this approach is that it
reduces some long-standing physical questions to standard problems in 
algebra and representation theory. Here are some examples:
\nxt As already discussed, we expect that the classification of CFTs with 
     given chiral data $\calc$ amounts
     to classifying Morita classes of symmetric special Frobenius algebras
     in the category $\calc$. In particular, physical modular invariant 
     partition functions of so-called \cite{mose2}
     automorphism type are classified by the Brauer group of $\calc$.
\nxt The classification of boundary conditions and defect lines 
     (that respect the underlying chiral symmetry) is reduced to the
     standard rep\-re\-sen\-tation theoretic problem of classifying modules
     and bimodules. As a consequence, powerful methods like induced modules
     and reciprocity theorems are at our disposal.
\nxt The problem of deforming CFTs is related to the problem of deforming
     algebras, which is a cohomological question. For the moment, the only
     known results in this direction are rigidity theorems \cite{etno}: a
     rational CFT cannot be deformed within the class of rational CFTs. 
     This is but one, though in our opinion not the least important, reason
     to make an effort to get a better understanding of non-rational 
     conformal field theory.

\smallskip

Indeed, in our opinion the biggest challenge in the theory is to go beyond the
rational case, i.e.\ admit categories that are not semi-simple any more
and which do not have such benign duality properties. Not too much is
known about such theories on a rigorous level. This is partly 
due to the lack of a class of well-understood examples. In the rational case,
the theory of Kac-Moody algebras -- or rather, quite specifically, of affine
Lie algebras -- has been of enormous help for the development of
conformal field theories. Similarly, the lack of knowledge concerning
non-rational conformal field theories can be regarded as reflecting
our comparatively poor understanding of the representation
theory of non-compact forms of Kac-Moody algebras of affine type.

\bigskip

\noindent
{\bf Acknowledgments}:\\
We are grateful to J\"urg Fr\"ohlich for helpful comments on the manuscript.

 \newcommand\wb{\,\linebreak[0]} \def\wB {$\,$\wb}
 \newcommand\Bi[1]   {\bibitem{#1}}
 \newcommand\J[5]   {{\sl #5}, {#1} {#2} ({#3}) {#4} }
 \newcommand\JK[4]  {{\sl #4}, {#1} {#3} }
 \newcommand\JKL[4]  {{#2} ({#3}) {#4} }
 \newcommand\K[6]    {\ {\sl #6}, {#1} {#2} ({#3}) {#4}}
 \newcommand\Prep[2] {{\sl #2}, preprint {#1}}
 \newcommand\BOOK[4] {{\em #1\/} ({#2}, {#3} {#4})}
 \newcommand\inBO[7] {{\sl #7}, in:\ {\em #1}, {#2}\ ({#3}, {#4} {#5}), p.\ {#6}}
 \newcommand\iNBO[6] {{\sl #6}, in:\ {\em #1}, {#2}\ ({#3}, {#4} {#5})}
 \def\jf    {J.\ Fuchs}
 \def\AMS   {{American Mathematical Society}}
 \def\aste  {Ast\'e\-ris\-que}
 \def\bams  {Bull.\wb Amer.\wb Math.\wb Soc.}
 \def\coma  {Con\-temp.\wb Math.}
 \def\comp  {Com\-mun.\wb Math.\wb Phys.}
 \def\cpma  {Com\-pos.\wb Math.}
 \def\duke  {Duke\wB Math.\wb J.}
 \def\fiic  {Fields\wB Institute\wB Commun.}
 \def\ijmp  {Int.\wb J.\wb Mod.\wb Phys.\ A}
 \def\imrn  {Int.\wb Math.\wb Res.\wb Notices}
 \def\jams  {J.\wb Amer.\wb Math.\wb Soc.}
 \def\jgap  {J.\wb Geom.\wB and\wB Phys.}
 \def\josp  {J.\wb Stat.\wb Phys.}
 \def\jram  {J.\wB rei\-ne\wB an\-gew.\wb Math.}                              
 \def\leni  {Lenin\-grad\wB Math.\wb J.}
 \def\maan  {Math.\wb Annal.}
 \def\mpla  {Mod.\wb Phys.\wb Lett.\ A}
 \def\nkwg  {Nachr.\wb K\"onigl.\wb Ges.\wB G\"ottingen, Math-phys.\wB Klasse}
 \def\nupb  {Nucl.\wb Phys.\ B}
 \def\pspm  {Proc.\wb Symp.\wB Pure\wB Math.}
 \def\sebo  {S\'emi\-naire\wB Bour\-baki} 
 \def\zfpc  {Zeitschr.\wb Phy\-sik C}
 \def\AP     {{Academic Press}}
 \def\NH     {{North Holland Publishing Company}}
 \def\PL     {{Plenum Press}}
 \def\SV     {{Sprin\-ger Ver\-lag}}
 \def\WS     {{World Scientific}}
 \def\Ad     {{Amsterdam}}
 \def\Be     {{Berlin}}
 \def\NY     {{New York}}
 \def\PR     {{Providence}}
 \def\Si     {{Singapore}}

\bibliographystyle{amsalpha}

\end{document}